\documentstyle [12pt, A4]{article}
\thicklines                     
\begin{document}
\rightline {CbNU-Th-94-27}
\rightline {June, 1994}

\vspace{.0in}
\begin{center}
{\large\bf Non-hermitian techniques of canonical transformations in quantum
mechanics.}\\[.2in]
{Haewon Lee\footnote{hwlee@cbucc.chungbuk.ac.kr}
and W. S. l'Yi\footnote{wslyi@cbucc.chungbuk.ac.kr}}\\[.15in]
{\it Department of Physics\\
Ch'ungbuk National University\\
Ch'ongju, 360-763, Ch'ungbuk, Korea} \\[.5in]
\end{center}

\begin{center}
{\bf ABSTRACT}\\
\end{center}
\begin{quotation}
The quantum mechanical version of the four kinds of classical canonical
transformations is
investigated by using non-hermitian operator techniques.
To help understand the usefulness of this appoach the eigenvalue problem of a
harmonic oscillator
is solved in two different types of canonical transformations.
The quantum form of the classical Hamiton-Jacobi theory is also employed to
solve time dependent
Schr\"odinger wave equations, showing that when one uses the classical action
as
a generating function of the quantum canonical transformation of time
evolutions of state vectors,
the corresponding propagator can easily be obtained.
\end{quotation}
\smallskip
PACS number(s): 03.65.-w, 03.65.Ca, 04.20.Fy, 04.60.Ds

\newpage

{\large\bf 1. Introduction}\vspace{.15in}

The idea of canonical transformations is one of the highlights of
classical mechanics\cite{classical_mechanics}.
It is not only theoretically but also practically important,
and provides some clue to the quantization of classical systems.
But the interesting point is that
even though the canonical transformations and the
Hamilton-Jacobi theory are very helpful for solving
classical equations of motions there appeared until now no serious
attempts to use these ideas while solving quantum mechanical problems.
One of the reasons may be that as long as one restricts himself to the unitary
forms
of canonical transformations there is no room for nontrivial transformations.

It was Anderson who seriously began to doubt the usual usage of the unitary
canonical
transformations, and he initiated a non-unitary technique of canonical
transformations\cite{anderson}.
His idea is that the commutation relations $[q_r, p_s]=i\delta_{rs},$
$[q_r,q_s]=[p_r,p_s]=0$
are preserved not only under unitary transformations but also following
similarity
transformations
\begin{equation}\label{similarlity_transformation}
Q_r=Cq_rC^{-1}, \;\;P_s=Cp_sC^{-1}.
\end{equation}
Observing the fact that any canonical transformation can be decomposed into
three basic
canonical transformations, he computed the eigenvalue equation of harmonic
oscillators and
also calculated propagators for some model cases.
Even though his idea is quite general it lacks clear classical counterparts.

In this paper we follow the traditional ``mixed matrix element technique'' of
canonical transformations\cite{transformation_form}.
But the novel difference is that our mixed matrix elements
are not unitary, thus allowing us to incorporate Anderson's idea of non-unitary
canonical
transformations.
In fact it is shown that the quantum versions of canonical transformations
exist and
formally are similar to the classical ones.   Using the full classical
properties of canonical
transformations we are able to solve the eigenvalue equation of a harmonic
oscillator in a
canonically-transformed new space.  Even the time-dependent Schr\"odinger
equations of free particles and harmonic oscillators can be solved by using the
quantum version
of the Hamilton-Jacobi theory.

This paper  is organized in the following way.  Non-unitary quantum canonical
transformations
which have classical analogies are introduced in chapter two.
In chapter three, the quantum version of a classical canonical transformation
is used
to solve the energy eigenstates of  harmonic oscillators.  The time evolutions
of
Schr\"odinger wave equations of the free particles and the harmonic ocillators
are also
solved by using the quantum version of the Hamilton-Jacobi theory.
Conclusions and further discussions are given in chapter four.

{\large\bf 2. Quantum canonical transformations.}\vspace{.15in}

In classical mechanics there are four different types of canonical
transformations, depending
on the forms of generating functions $F_1(q_r, Q_s,t),$ $F_2(q_r, P_s,t),$
$F_3(p_r, Q_s,t),$
and $F_4(p_r, P_s,t).$  Even though some of them are related there are
transformations
which cannot be described by any other type of transformation.
In this paper the quantum versions of the first and second types of canonical
transformations are presented. The other two remaining ones can be inferred
from these.

{\bf a) Canonical transformations of the first kind.}

Suppose $|q'\rangle=|q'_1,\dots,q_f\rangle$ is a simultaneous eigenket of
observables $q_r,$ $r=1, \dots, f,$ such that
\begin{eqnarray}
 q_r|q'\rangle &=& q'_r|q'\rangle, \\ \nonumber
 \langle q' | q'' \rangle  &=& {1\over\rho(q')} \delta(q'-q''),\\ \nonumber
 \int d^fq'\, |q'\rangle \rho(q') \langle q'| &=& 1.
\end{eqnarray}
{}From now on we use the convention that various eigenvalues of an observable
$q_r$ are denoted by
attaching primes such as $q'_r,$ $q''_r,$ etc.
Schr\"odinger equations are sometimes readily solvable in  different basis kets
$|Q't\rangle=|Q'_1, \dots, Q'_f,t\rangle$
which are defined with respect to $|q'\rangle$ by
\begin{equation}\label{definition}
\langle q' | Q't \rangle = e^{\textstyle{iF(q'_r,Q'_s,t)}},
\end{equation}
where $F$ is a function of real numbers $q'_r$ and $Q'_s,$ and time $t.$
Transformations of
these kinds were investigated from the early days of quantum
mechanics\cite{transformation_form}.
One should note that for an arbitrary
function $F$ the completeness condition of $|Q't\rangle$ is not guaranteed.
Transformations which do not satisfy the completeness condition lose some
information.
Here we consider only the generating functions which meet the completeness
condition. General case will considered in appendix B.  Suppose that for some
density function
$\rho(Q',t)$ they are complete, that is
\begin{eqnarray}\label{completeness}
\langle Q't | Q''t \rangle &=& {1\over\rho(Q',t)}\delta(Q'-Q''),\\ \nonumber
\int d^fQ'\, |Q't\rangle \rho(Q',t) \langle Q't | &=& 1.
\end{eqnarray}
In this case we are able to define other observables $Q_r, r=1, \dots, f,$ such
as
\begin{equation}\label{eigen_val_Q}
Q_r|Q't\rangle=Q'_r|Q't\rangle.
\end{equation}

To get the quantum analogy of classical canonical transformations let us assume
that
$F(q_r,Q_s,t)$ is a ``well-ordered'' operator in the sense that
it is a sum  of $q$-functions multiplied by $Q$-functions on the right.
In that case we have
\begin{equation}
\langle q' | F(q_r,Q_s,t) |Q't\rangle =F(q'_r,Q'_s,t)\langle q'|Q't\rangle.
\end{equation}
In this case ``the non-hermitian canonical momentum operators'' $p_r$ and
$P_r,$ which are
defined by
\begin{eqnarray}
\langle q'|p_r|Q't\rangle &=& -i{\partial\over \partial q'_r} \langle
q'|Q't\rangle, \\
\langle Q't|P_r|q'\rangle &=& -i{\partial\over \partial Q'_r} \langle
Q't|q'\rangle,
\end{eqnarray}
can be recast in the familiar forms of classical mechanics
\begin{equation}\label{p_r}
p_r={\partial F\over \partial q_r},\;\;
P_r=-{\partial F^{\dagger}\over\partial Q_r}.\label{P_r}
\end{equation}
Note that even if $q_r$ and $Q_s$ are observables the corresponding canonical
momentum
operators may not be hermitian. In fact their hermitiain conjugations are
\begin{equation} \label{p_dag}
p_r^{\dagger}=\rho(q)^{-1}p_r\rho(q), \;\;
P_r^{\dagger}=\rho(Q,t)^{-1}P_r\rho(Q,t).
\end{equation}
These non-hermitian properties are essential for our investigation and will be
discussed
more carefully in the last part of this section.

We now proceed to get the Schr\"odinger equation of motion in the
canonically-transformed
$Q$-space. Let $|t\rangle$ be a Schr\"odinger ket whose
motion is given by $i{d \over dt}|t\rangle = H(q_r,p_s,t) |t\rangle.$
The $Q$-space wave function $\phi(Q'_r,t)=\langle Q't|t\rangle$ in terms of
$q$-space wave
function $\psi(q',t)=\langle q'|t\rangle$ is given by
\begin{equation}
\phi(Q'_r,t)=\int e^{ -iF^{*}(q'_r, Q'_s,t) } \psi(q'_r,t) \rho(q'_r) d^f q'.
\end{equation}
Applying the usual wave equation to $\psi(q'_r,t)$
one obtains
\begin{equation}\label{Q_space_wave_equation}
i{\partial \over \partial t} \phi(Q'_r,t)
	= K(Q'_r, -i{\partial \over \partial Q'_s}, t) \phi(Q'_r,t),
\end{equation}
where the $Q$-space Hamiltonian $K(Q_r, P_s, t)$ is
\begin{equation}\label{K}
K(Q_r,P_s,t)=H(q_r, p_s,t) + {\partial F^{\dagger} \over \partial t}.
\end{equation}
After solving (\ref{Q_space_wave_equation}) the true wave function in $q$-space
is constructed by
\begin{equation}
\psi(q'_r,t)=\int e^{ iF(q'_r, Q'_s,t) } \phi(Q'_r,t) \rho(Q'_r,t) d^f Q'.
\end{equation}

At this point we would like to emphasize that to get the completeness condition
(\ref{completeness_2}), oftentimes one is forced to use ``wrong'' density
functions in
the original $q$-space. For example, in a one-dimensional case the proper
density function $\rho(x)$ of
the cartesian coordinate $x$ is constant. But when one uses a generating
function of the form
$F(x,Q)=x^3 Q,$ he is obliged to employ the wrong density function
$\widetilde{\rho}(x)=2x^3.$
Even more, because of (\ref{p_dag}), the hamiltonian becomes non-hermitian.
These two problems cancel each other out to resolve the dilemma.  This can be
seen in the following way.
First rescale the original basis kets such as
\begin{equation}
\widetilde{|q'\rangle}
=|q'\rangle S(q')^{-1}, \label{newket}
\end{equation}
where
\begin{equation}
 S(q) = \left({\widetilde{\rho} (q) \over \rho(q)}\right)^{{1 \over 2}},
\end{equation}
and $\widetilde{\rho}(q)$ is a new density function.
These rescaled kets satisfy the following completeness conditions
\begin{eqnarray}
\widetilde{\langle q'} | \widetilde{q'' \rangle}
&=& {1 \over \widetilde{\rho}(q')} \delta(q'-q''),\\ \nonumber
\int d^f q' \widetilde{|q'\rangle} \widetilde{\rho}(q')\widetilde{\langle q' |}
&=& 1.
\end{eqnarray}
The corresponding conjugate momentum operator $\widetilde{p}_r$ defined by
\begin{equation}
\widetilde{\langle q' |} \widetilde{p}_r\widetilde{|q''\rangle}
= -i {\partial \over \partial q'_r} \widetilde{\langle q'|}
\widetilde{q''\rangle}
\end{equation}
is related to $p_r$ by a similarity transformation
\begin{equation}
\widetilde{p}_r=Sp_rS^{-1}.
\end{equation}

To investigate how the Schr\"odinger wave equation changes under this
similarity transformation
we make use of the standard ket $\rangle$ which is introduced by Dirac in his
famous book
on quantum mechanics\cite{dirac_book}.
For a given set of basis kets $|q'\rangle$ it, by definition, satisfies
\begin{equation}
\langle q'|\rangle=1.
\end{equation}
The reason why one introduces the standard ket is that
any state $|1\rangle$ such as $\langle q'|1\rangle = \psi(q')$ can be
written as $\psi(q)\rangle,$ that is, a function $\psi(q)$ of an observable $q$
operating on
the standard ket $\rangle.$   When one chooses another set of basis kets
$\widetilde{|q'\rangle}$
he may use another standard ket $\widetilde{\rangle}$ such as
$\widetilde{\langle q'|}\widetilde{\rangle}=1.$
{}From (\ref{newket}) it is clear that
\begin{equation}
\widetilde{\rangle} = S(q) \, \rangle.
\end{equation}
Applying this similarity transformation to the original Schr\"{o}dinger
equation
\begin{equation}
H(q_r, p_s,t)\,\psi(q_r,t)\,\rangle =i{\partial\psi \over \partial t}\,\rangle
\label{hermitian_equation}\end{equation}
the following similar non-hermitian Schr\"odinger equation is produced
\begin{equation}
H(q_r, \widetilde{p}_s,t)\,\psi(q_r,t)\,\widetilde{\rangle}
=i{\partial\psi \over \partial
t}\,\widetilde{\rangle}.\label{non_hermitian_equation}
\end{equation}
It means that as far as one uses a $p_r=-i{\partial \over \partial q_r}$
representation
he may freely choose different density functions while the wave function and
the Schr\"odinger equation remain intact.  The sole requirement is that one
should use
non-hermitian hamiltonian, which is tolerable as far as the wave function is
concerned.
But the physically meaningful transition probabilities should be constructed by
using the original density function at which the hamiltonian is hermitian, and
so  also for
the wave function normalization.
This fact will
be explained when the quantization of the harmonic oscillator is treated.

{\bf b) Canonical transformations of the second type.}

Consider two observables $q_r$ and $P_s$ with corresponding eigenkets
$|q'\rangle$ and
$|P't\rangle$ which are related by
\begin{equation}
\langle q'|P't\rangle=e^{\textstyle{iF(q'_r,P'_s,t)}}
\end{equation}
where the generating function $F(q'_r,P'_s,t)$ is a function of real variables
$q'_r,$ $P'_s,$ and $t.$
Since $P_s'$ is a continuous number, the corresponding ket is normalized by
\begin{equation}
\langle P't|P''t\rangle={1 \over \rho(P',t)} \delta(P'-P'').
\end{equation}
Consider a ``well-ordered operator'' $F(q,P,t)$ such that
\begin{equation}
\langle q'|F(q_r,P_s,t)|P't\rangle=F(q'_r,P'_s,t)\langle q'|P't\rangle.
\end{equation}
The operators defined by
\begin{equation}
\langle q'|p_r|P't\rangle=-i{\partial\over\partial q'_r} \langle
q'|P't\rangle,\;\;\;
\langle P't|Q_r|q'\rangle=i{\partial\over\partial P'_r} \langle P't|q'\rangle
\end{equation}
can be written as
\begin{equation}
	p_r={\partial F\over\partial q_r},\;\;\;
	Q_r={\partial F^{\dagger}\over\partial P_r}.
\end{equation}
The wave equation in $P$-space is
\begin{equation}
i{\partial \over \partial t}\phi(P'_r,t)
        =K(i{\partial \over \partial P'_r}, P'_s,t)\phi(P'_r,t),
\end{equation}
where $\phi(P'_r,t)=\langle P't|t\rangle$ and
$K=H+{\partial F^{\dagger} \over \partial t}$ is a hamiltoinan in $P$-space.
\newpage
{\large\bf 3. Applications of quantum canonical transformations.}\vspace{.15in}

As much as the concept of classical canonical transformations is practically
helpful for solving
equations of motion, so is also the technique of the quantum canonical
transformations.
It can be used not only for eigenvalue problems, but also
for the time evolutions of state vectors including the time-dependent
perturbations.
In this paper only some ideal physical systems are treated to clarify this
rather new concept.

{\bf a) Energy eigenfunctions of a harmonic oscillator.}

As an application of type one canonical transformations, consider the following
hamiltonian of a
harmonic oscillator
\begin{equation}\label{HO}
H(q,p)={1\over2}\left(q^2+p^2\right).
\end{equation}
To solve the energy eigenvalue equation one may choose a generating function
$F(q,Q)={1\over 2}q^2\cot Q$ which is well known
from classical mechanics. In this case $|Q'\rangle$ is defined by
\begin{equation}\label{gen_of_HO_1}
\langle q'|Q' \rangle =e^{i{1\over 2}q'^2\cot Q'}.
\end{equation}
Since $F$ is an even function of $q,$ any wave function in $Q$-space becomes an
even function in
$q$-space whenever $Q$-space is transformed to $q$-space by using the
transition
amplitute (\ref{gen_of_HO_1}).
In fact one may prove that
\begin{equation}\label{completenetss_of_Q}
\int^\pi_0 dQ' |Q'\rangle \rho(Q') \langle Q'| = {\bf 1_+},
\end{equation}
where $\rho(Q)=\csc^2Q,$ and ${\bf 1_+}$ is the parity even projection operator
in $Q$-space.
One may push along this line only to obtain parity even wave functions. The
quantization
of energy is very intuitive when a  generating function of this type is
used.  (See appendix A for more details.)

To get the complete wave functions the following definition of $Q$ is employed
\begin{equation} \label{Q=p+iq}
Q=p+iq.
\end{equation}
Assuming that it is a type one canonical transformation one can write $p$ as
$p={\partial F(q,Q) \over \partial q}.$  Plugging this into (\ref{Q=p+iq}),
$F(q,Q)$ can be solved to give
\begin{equation}\label{F_OF_HO}
F(q,Q)=-{i\over 2} q^2 + qQ.
\end{equation}
Since this generating function contains a bilinear term in $q$ and $Q,$ there
would be no such
problem as the one which we encountered previously.
By direct calculations it can be shown that basis kets $|Q'\rangle$ are in fact
complete, and
the corresponding density functions are
\begin{equation}\label{density_of_HO}
\rho(q)={1\over 2\pi} e^{ -q^2}, \;\; \rho(Q)=1.
\end{equation}
The transformed operator $P$ which can be read off from (\ref{p_r}) and
(\ref{F_OF_HO})
are
\begin{equation}
P=-q.
\end{equation}
The hamiltonian $K$ in $Q$-space is
\begin{equation}\label{K_OF_HO}
K=iQP+{1\over 2} Q^2 +{1\over 2}.
\end{equation}
This hamiltonian is non-hermitian.
Even worse, since $p=Q-iq$ is non-hermitian, the original hamiltonian
(\ref{HO}) is non-hermitian.
All these unsual facts are reflected in the ``wrong density functions''
given by (\ref{density_of_HO}).  As we already pointed out in the previous
chapter
there would be no problem at all
as far as the wave function is concerned.  One may convince himself by checking
the final
wave function (\ref{wave_function_of_HO}).

By (\ref{K_OF_HO}) the time-independent Schr\"odinger wave equation in
$Q$-space is
\begin{equation}
\left( Q{\partial \over \partial Q}+{1\over 2} Q^2 +{1\over 2}
\right)\phi(Q)=E\,\phi(Q)
\end{equation}
This first order differential equation can easily be solved giving the
following wave function
\begin{equation}
\phi(Q)=N\, Q^\nu\,e^{ -{1\over 4}Q^2},
\end{equation}
where $\nu=E-{1\over 2},$ and $N$ is a normalization constant which must be
determined after
the wave function in $q$-space
\begin{equation}\label{psi_of_HO}
\psi(q)=N e^{ {1\over 2}q^2 } \int^{\infty}_{-\infty}
	e^{  iqQ -{1\over 4} Q^2 } Q^\nu dQ
\end{equation}
is evaluated.  It is not difficult to show that this is proportional to
$D_\nu(\sqrt{2}q),$
where $D_\nu(z)$ is the {\it parabolic cylindrical function} defined
by\cite{parabolic}
\begin{equation}
D_\nu(z)=\sqrt{ {2\over \pi} } 2^\nu e^{ -{\pi\over2} \nu i} e^{ {z^2\over 4} }
	\int^{\infty}_{-\infty} x^\nu e^{-2x^2 + 2ixz} dx, \;\; {\rm Re}\,\nu > -1.
\end{equation}
For noninteger $\nu,$ it is known that $D_{\nu}(z)$ diverges as $z$ goes to
$-\infty$\cite{D(z)}.
That means the energy of the hamonic oscillator should be
\begin{equation}
E=n+{1\over 2},\;\; n=0, 1, \dots .
\end{equation}
The corresponding wave function is
\begin{eqnarray}\label{wave_function_of_HO}
\psi_n(q) &=& N e^{ {1\over 2}q^2 } \left( {1\over i} {d\over dq} \right)^n
	\int^{\infty}_{-\infty} e^{  iqQ -{1\over 4} Q^2 } dQ \\ \nonumber
	&=& N' (-)^n e^{ {1\over 2}q^2 } ({d\over dq})^n e^{-q^2}.
\end{eqnarray}
This wave function coincides with the usual form constructed directly in the
cartesian
coordinate system.

{\bf b) Propagators}

Our new technique can also be used to solve the time dependent Schr\"odinger
wave equations.
In classical mechanics there is an elegant way of solving equations of motion
known as
the Hamilton-Jacobi theory, which uses the classical action as the generating
function.
In that case the transformed hamiltonian vanishes. Since the transformed
quantum hamiltonian
(\ref{K}) looks like the classical one, one may expect similar results for
quantum mechanics.
But because of the noncommutative properties of quantum canonical variables,
there may arise terms which are proportional to $\hbar$ which we simply put
equal to one.

To convince ourselves the advantage of our quantum Hamilton-Jacobi theory for
solving the time-dependent Schr\"odinger equations, we first consider
one-dimensional free
particles.  The corresponding hamiltonian is
\begin{equation}
H={1\over 2}p^2.
\end{equation}
The classical action for $Q$ to $q$ time-evolution during $t$ is
\begin{equation}
S(q,Q,t)={(q-Q)^2\over 2t}.
\end{equation}
That means the best well-ordered quantum generator for this time evolution is
\begin{equation}
F(q,Q,t)={1\over 2t}(q^2-2qQ+Q^2).
\end{equation}
Using this generating function the transformed ket $|Q't\rangle$ is defined by
\begin{equation}
\langle q'|Q't\rangle = e^{iF(q',Q',t)}.
\end{equation}
Note that we are using Schr\"odinger picture. That is the reason why the
original basis ket
$|q'\rangle$ is time independent.
By the direct calculation we obtain the following density functions
\begin{equation}
\rho(q)=1, \;\; \rho(Q,t)={1 \over 2\pi t}
\end{equation}
The canonical momentum $p,$ $P,$ and the $P$-space hamiltonian $K$
are given by
\begin{equation}
p = {q-Q \over t}, \;\; P=p, \;\; K = {i \over 2t}.
\end{equation}
Classical hamiltonian in $Q$-space vanishes, but in quantum physics it is
propotional to
$\hbar.$  The equation of motion in $Q$-space is
\begin{equation}
i{\partial \phi \over \partial t} = {i\over 2t} \phi,
\end{equation}
and the solution with a convenient normalization constant is
\begin{equation}
\phi(Q,t)= \sqrt{ {2\pi t \over i}}\phi(Q).
\end{equation}
The true wave equation in $q$-space is therefore
\begin{equation}\label {psi_in_phi}
\psi(q,t)=\int^{\infty}_{-\infty}
	{1 \over \sqrt{ 2 \pi it} } e^{i{ (q-Q)^2 \over 2t} } \phi(Q)dQ.
\end{equation}

To understand the physical meaning of this propagation equation
we investigate $t\rightarrow 0$ limit.
With the help of
\begin{equation}\label{delta_function}
\lim_{t\rightarrow 0} { e^{ i{(x'-x'')^2 \over 2t} } \over \sqrt{2\pi it}}
=\delta(x'-x'')
\end{equation}
it is clear that $\phi(q)$ is nothing but $\psi(q,0).$
It means that the free particle propagator is, by (\ref{psi_in_phi}),
\begin{equation}
G(q',t|q'',0)={1 \over \sqrt{ 2 \pi it} } e^{i{ (q'-q'')^2 \over 2t} }.
\end{equation}
At this point we would like to emphasize that by a direct path integral it is
known that
the general form of a propagator is\cite{feynmann_hibbs}
\begin{equation}
G(q',t|q'',0)= f(t) e^{iS_{cl} },
\end{equation}
where $S_{cl}$ is the classical action and $f(t)$ is an undetermined function
of time.
In our approach $f(t)$ is related to the density function $\rho(Q,t)$ and the
$\hbar$
proportional hamiltonian $K$ in $Q$-space.

Using a similar technique one can solve the time-dependent Schr\"odinger
equation for harmonic
osillators.  The well-ordered generating function would be
\begin{equation}\label{F_of_HO}
F(q,Q,t)={1\over 2} (q^2 +Q^2)\cot t -qQ\csc t
\end{equation}
which is just the classical action for $Q$ to $q$ time-evolution during
$t$\cite{classical_mechanics}.
One can prove the completeness condition for $|Q',t\rangle,$ obtaining
simultaneously the
following density functions
\begin{equation}\label{density_HO}
\rho(q)=1,\;\; \rho(Q,t)={1\over 2\pi \sin t}.
\end{equation}
If $q$ and $Q$ commute, the tranformed hamiltonian $K$ vanishes.  But for
quantum theory
it does not but is proportional to $\hbar,$ that is,
\begin{equation}
K(Q,P,t)={i\over 2} \cot t.
\end{equation}
Using this hamiltonian the Schr\"odinger equation in $Q$-space can be solved
giving the following time-dependent wave function,
\begin{equation}
\phi(Q,t)=\sqrt{ {2\pi \sin t \over i}}\phi(Q).
\end{equation}
With the help of (\ref{density_HO}) and (\ref{delta_function})
one obtains following propagator of a harmonic oscillator
\begin{equation}\label{kernel_OF_HO}
G(q',t|q'',0)={1 \over \sqrt{ 2 \pi i\sin t} } e^{iF(q',q'',t)},
\end{equation}
where $F(q',q'',t)$ is the classical action given by (\ref{F_of_HO}).

To see why the quantum generating function which has the classical analogy is
so successful,
consider the following classical generating function for infinitesmal time
evolution
\begin{equation}\label{infinitesmal_CT}
F(q_r, P_s) = \sum_r q_r P_r -\delta t H(q_r, P_s),
\end{equation}
where $H= {1\over 2} \sum_r P^2_r  + V(q_s)$ is the usual hamiltonian.
Using this we define a state $|P'\rangle$ by $\langle q'|P'\rangle= e^{
iF(q',P') }.$
Then the density functions both in $q$ and $P$ spaces are trivial and $Q_r,$
which is defined
by ${\partial F^{\dagger} \over \partial P_r},$ is hermitian.   It means that
all the eigenkets
$|Q'\rangle$ form a complete set. To get the physical meaning of $|Q'\rangle$
consider
$\langle q'|Q'\rangle.$
Using the completeness of eigenkets $|P'\rangle$ one has
\begin{eqnarray}\label{Qq}
\langle q'|Q'\rangle
	&=& \int^{\infty}_{-\infty} {dP' \over 2\pi} \langle q'|P'\rangle \langle
P'|Q'\rangle\\
		\nonumber
	&=& \int^{\infty}_{-\infty} {dP' \over 2\pi} e^{i\sum q'_r P'_r - i\delta t
H(q'_r,P'_s)}
		\langle P'|Q'\rangle .
\end{eqnarray}
Now the term $e^{i\sum q'_r P'_r}$ in the last part of this equation can be
written as
$\langle Q' \leftarrow q' | P'\rangle,$ where $|Q' \leftarrow q'\rangle$ is an
eigenket of $Q_r$
whose eigenvalue is $q'_r.$ Using this fact (\ref{Qq}) can be simplified as
\begin{equation}
\langle q'|Q'\rangle = \langle Q' \leftarrow q'|e^{-i\delta t H(Q_r,P_s)}
|Q'\rangle,
\end{equation}
that is $|Q' \leftarrow q'\rangle = e^{-i\delta t H} |q'\rangle.$
This mean that (\ref{infinitesmal_CT}) is a both classically and quantum
mechanically
correct generating function of the infinitesmal time evolutions.

{\large\bf 4. Conclusion and discussion}\vspace{.15in}

We would like to emphasize that when the quantum analogies of classical
canonical transformations
are seriously employed, one obtains useful quantum canonical transformations
which can be
used to solve either time-independent or time-dependent
Schr\"odinger equations.  Furthermore the generating operators of quantum
canonical
transformations can be inferred from the classical generating functions.
In this way the classical
canonical transformations have some part in quantum mechanics.
This classical analogy
is our strong point which becomes rather obscure when one uses abstract
similarity
transformation formalism. (See appendix B for more details.)
We expect that our idea will produce more fruitful results when applied to the
pertubation
theory of
quantum mechanics.

{\large\bf Acknowledgments}\vspace{.15in}

This is supported by the Basic Science Research Institute Program of the
Ministry of Education,
Korea, 1993.
\newpage
{\large \bf Appendix}\vspace{.15in}
\renewcommand{\theequation}{A.\arabic{equation}}
\setcounter{equation}{0}

{\bf A. Energy quantizations of harmonic oscillators}

Consider a harmonic oscillator whose hamiltonian is given by
\begin{equation}
H(q,p)={1\over 2}(q^2 + p^2).
\end{equation}
To get some insight into the energy quantization of a harmonic oscillator we
introduce the
following transformation
\begin{equation}\label{gen_of_HO_1_A}
\langle q'|Q' \rangle =e^{i{1\over 2}q'^2\cot Q'}.
\end{equation}
Because of (\ref{completenetss_of_Q}) this transformation is complete only for
the
parity even subspace of $\psi(q).$
The corresponding non-hermitian operators $p$ and $P$ are given by
\begin{eqnarray}
p&=&{\partial F\over \partial q}=q\cot Q\\ \nonumber
P&=&-{\partial F^{\dagger} \over \partial Q}={1\over 2}\csc^2Q q^2
\end{eqnarray}
{}From the commutation relation $[q,p]=i$ we have the following equation
\begin{equation}
[q,\cot Q]=iq^{-1}.
\end{equation}
Using this relation the $Q$-space hamiltonian $K(Q,P)=H(q,p)$ can be shown to
be
\begin{equation}
K(Q,P)=P+{3\over2}i\cot Q.
\end{equation}
The eigenvalue equation
$K(Q,P)\varphi=E\varphi$ in $Q$-space is therefore
\begin{equation}
-i\left({\partial\over\partial Q}-{3\over 2}\cot Q\right)\varphi=E\varphi.
\end{equation}
It can immediately be solved  giving
\begin{equation}\label{phi_in_Q}
\varphi(Q)=N\sin^{3\over 2}Qe^{iEQ},
\end{equation}
where $N$ is a normalization constant. Since $|Q'\rangle$ is defined by
(\ref{gen_of_HO_1_A})
the wave function in $Q$-space should satisfy $\varphi(Q'+\pi)=\varphi(Q').$
This means that
\begin{equation}
e^{ i({3\over 2} + E)\pi }=1.
\end{equation}
This together with the positive energy condition yields
\begin{equation}
E=(2n+{1\over 2}),\; n=0,1,2,\dots .
\end{equation}
The wave function $\psi(q')$ in $q$ space is given by
\begin{eqnarray}
\psi_{2n}(q')&=&N'\int^\pi_0 dQ' \rho(Q')\langle q'|Q'\rangle\varphi(Q')\\
\nonumber
	&=&N''\int^\pi_0 dQ' {1\over \sin^{1\over 2}Q'}
	\exp\left( {i\over 2}q'^2 \cot Q'+ i(2n+{1\over 2}) Q'\right).
\end{eqnarray}
When we compare this equation with the well known result for a harmonic
oscillator (\ref{kernel_OF_HO})
\begin{equation}
\sum_{n=0}^\infty\psi_{2n}(q')\psi^*_{2n}(0)e^{-i(2n+{1\over 2})T}
=\left({1\over2\pi i}\right)^{1\over2}
	{e^{{i\over 2}q'^2 \cot T} \over \sin^{1\over2}T},
\end{equation}
we obtain the correct parity even wave functions.

{\bf B. Generalizations and relation to similarity transformation approach
}\vspace{.15in}
\renewcommand{\theequation}{B.\arabic{equation}}
\setcounter{equation}{0}

In appendix B we generalize some results obtained in chapter two and
investigate the relation between our method and the similarity
transformation method given by (\ref{similarlity_transformation}).
For simplicity we consider only a one-dimensional case.

Suppose $|q'\rangle$ be a set of ket vectors in a Hilbert space and
$\widetilde{|q'\rangle}$ a new set of ket vectors in the same
Hilbert space. In general $\widetilde{|q'\rangle}$ may not form a complete
orthogonal set. One can make $\widetilde{|q'\rangle}$ complete by introducing
a new inner product in the same Hilbert space, which will be shown later
in this appendix.  In this case the original $|q'\rangle$
and the new $\widetilde{|q'\rangle}$ will be complete with respect to two
different
inner products, that is, with respect to the old and to the new ones,
respectively.
Then the dual vector $\widetilde{\langle q'|}$ should be defined using the new
inner
product. Suppose that $|q'\rangle$ and $\widetilde{|q'\rangle}$ are
complete with density functions $\rho(q')$ and $\widetilde{\rho}(q'),$
respectively.
Then one get
\begin{eqnarray}\label{completeness_2}
 \int dq'\, |q'\rangle \rho(q') \langle q'| &=& 1, \\ \nonumber
 \int dq'\, \widetilde{|q'\rangle} \widetilde{\rho}(q')
                \widetilde{\langle q'|} &=& 1.
\end{eqnarray}
For some cases $|q'\rangle$ or $\widetilde{|q'\rangle}$ may be incomplete.
In that case  the left hand side of (\ref{completeness}) should be replaced by
appropriate
projection operators. The operators
$q$ and $\widetilde{q}$ which are defined by
\begin{equation}
  q|q'\rangle = q'|q'\rangle, \;\;
    \widetilde{q}\widetilde{|q'\rangle} = q'\widetilde{|q'\rangle}
\end{equation}
will be hermitian with respect to the old inner product and the new one,
respectively.
It can be shown that the two density functions together with
$\widetilde{|q'\rangle}$ determine
the new inner product uniquely. To clarify this point, let us consider the
followings.

Let us write $\widetilde{|q'\rangle} \equiv C(t)|q'\rangle,$ where
$C(t)$ can be regarded as a time-dependent similarity transformation.
Then $\widetilde{q}=CqC^{-1}$. A new dual vector $\widetilde{\langle q'|}$ can
be expressed in terms of the old one as
\begin{equation}
    \widetilde{\langle q'|} = \langle q'|C^{\dagger}M ,
\end{equation}
where $M$ is an operator defining the new inner product. In other words,
$(\alpha,\beta)=\langle\alpha|M|\beta\rangle,$ where $(\, , \, )$ denotes the
new inner product. Using the completeness relation we can find
\begin{equation}\label{simil1}
    C^{\dagger}MC=\frac{\rho(q)}{\widetilde{\rho}(q)},
\end{equation}
and conclude that $\rho,$ $\widetilde{\rho}$ and $C$ determine $M$ completely.
In chapter two we specialized ourselves to the cases where both of
$|q'\rangle$ and $\widetilde{|q'\rangle}$ form  complete orthogonal
sets with respect to the same inner product. In this case, $M=1$ and $C$ is
``almost unitary'' in the sense that $C\sqrt{\frac{\widetilde{\rho}}{\rho}}$ is
unitary.  This is not always true. As shown above, in general
cases we are forced to introduce two different inner products.

Now we study how the momentum operators transform. Using (\ref{simil1}) and
the fact that momentum operators are defined by
\begin{equation}
    \langle q'|p=-i\frac{\partial}{\partial q'}\langle q'|, \;\;
    \widetilde{\langle q'|} \widetilde{p} = -i\frac{\partial}{\partial
q'}\widetilde{\langle q'|}  ,
\end{equation}
one can easily prove that
\begin{equation}
    \widetilde{p} = \widetilde{C} p \widetilde{C}^{-1},\;\;
        \widetilde{q} = \widetilde{C} q \widetilde{C}^{-1}  ,  \label{simil2}
\end{equation}
with $\widetilde{C} = (C^\dagger M)^{-1}.$
Next we consider the Schr\"odinger equation
\begin{equation}
    H(q,p,t)|t\rangle = i\frac{d}{dt} |t\rangle .
\end{equation}
Multiplying both sides by $\widetilde{\langle q'|},$  we have
\begin{equation}
\widetilde{\langle q'|} H(q,p,t) |t\rangle
    = i {\partial  \over \partial t} \widetilde{\langle q'|} t\rangle
    - \widetilde{\langle q'|} \dot{G}| t\rangle ,
\end{equation}
where $\dot{G}$ is defined by
\begin{equation}
    \widetilde{\langle q'|}\dot{G}
        = i {\partial  \over \partial t} \widetilde{\langle q'|}.
\end{equation}
One can express $\dot{G}$ using the similarity tranformation,
\begin{equation}
    \dot{G} = i \widetilde{C} {\partial \over \partial t} \widetilde{C}^{-1} .
\end{equation}
Denoting
\begin{equation}
    K(\widetilde{q},\widetilde{p},t) \equiv H(q,p,t) + \dot{G},
\end{equation}
one can write the Schr\"odinger equation in $\widetilde{q}$-space as
\begin{equation}
    K(q',-i{\partial \over \partial q'},t) \widetilde{\langle q'|} t\rangle
    = i {\partial  \over \partial t} \widetilde{\langle q'|} t\rangle
\end{equation}

To find $K(\widetilde{q},\widetilde{p},t)$ in terms of $\widetilde{q}$ and
$\widetilde{p},$
(\ref{simil2}) may, in principle, be used.
However as in chapter two, some interesting results can be obtained
if the transformation is expressed in the form
\begin{equation}
    \langle q' \widetilde{|q''\rangle} = e^{iF(q',q'',t)} , \;\;
    \widetilde{\langle q''|}q'\rangle = e^{-iG(q'',q',t)} ,
\end{equation}
where $F$ and $G$ are some functions obeying the relation
\begin{equation}
    \int dq'' e^{iF(q',q'',t)} \widetilde{\rho}(q'',t) e^{-iG(q'',q''',t)}
        = {\delta(q'-q''') \over \rho(q')}  . \label{simil4}
\end{equation}
As in chapter two we assume that $F(q,\widetilde{q},t)$ and
$G(\widetilde{q},q,t)$
are ``well-ordered'' operators in the sense that in $F(q,\widetilde{q},t)$ all
$q$'s
are on the left, and in $G(\widetilde{q},q,t)$ all q's are on the right.
Then (\ref{P_r}) corresponds to
\begin{equation}
    p={\partial \over \partial q} F(q,\widetilde{q},t) ,\;\;
    \widetilde{p} = -{\partial \over \partial \widetilde{q}}
G(\widetilde{q},q,t),\label{simil3}
\end{equation}
and we also have
\begin{equation}
    \dot{G} = {\partial \over \partial t} G(\widetilde{q},q,t)  .
\end{equation}
Eq.(\ref{simil3}) can be used to get $\widetilde{q}$ and $\widetilde{p}$ in
terms of $q$ and $p.$
In ``almost unitary'' cases considered in chapter two, we get
\begin{equation}
    G(\widetilde{q},q,t)=F(q,\widetilde{q},t)^\dagger.
\end{equation}
The two density functions can be absorbed into $F$ and $G,$  which is obvious
from (\ref{simil4}-16). In this case we have $\rho=1$ and $\widetilde{\rho}=1$.

\newpage

\newcommand{\pra}[3]{ Phys. Rev. A {\bf #1}, #2 (#3)}
\newcommand{\pl}[3]{ Phys. Lett. B {\bf #1}, #2 (#3)}
\newcommand{\np}[3]{ Nucl. Phys. {\bf B#1}, #2 (#3)}
\newcommand{\mpl}[3]{ Mod. Phys. Lett. A {\bf #1}, #2 (#3)}
\newcommand{\same}[3]{ {\bf #1}, #2 (#3)}
\newcommand{\book}[2]{ {\it #1} (#2) }

\end{document}